\documentclass[12pt]{article}
\usepackage{epsfig}
\newcommand{\beq}{\begin{equation}}
\newcommand{\eeq}{\end{equation}}
\newcommand{\beqa}{\begin{eqnarray}}
\newcommand{\eeqa}{\end{eqnarray}}
\newcommand{\beqar}{\begin{eqnarray*}}
\newcommand{\eeqar}{\end{eqnarray*}}
\newcommand{\eg}{{\it e.g.,}\ }
\newcommand{\ie}{{\it i.e.,}\ }
\newcommand{\labell}[1]{\label{#1}} 
\newcommand{\reef}[1]{(\ref{#1})}

\begin{document}
\addtolength{\baselineskip}{1.5mm} 

\thispagestyle{empty}

\hfill{}

\hfill{}

\hfill{CERN-TH/2001-113}

\hfill{hep-th/0104206}

\vspace{32pt}

\begin{center}
\textbf{\Large Macroscopic and Microscopic Description of Black
Diholes}\\

\vspace{48pt}

Roberto Emparan$^a$\footnote{roberto.emparan@cern.ch. Also at
Departamento de F{\'\i}sica Te\'orica, Universidad del Pa{\'\i}s Vasco,
Bilbao, Spain.} and Edward Teo$^b$

\vspace{12pt}

\textit{
$^a$Theory Division, CERN, CH-1211 Geneva 23, Switzerland\\
$^b$Department of Physics, National University of Singapore, Singapore
119260
}

\end{center}
\vspace{48pt}

\begin{abstract} 
\addtolength{\baselineskip}{1.2mm} 

We study configurations consisting of a pair of non-extremal black holes
in four dimensions, both with the same mass, and with charges of the
same magnitude but opposite sign---diholes, for short. We present such
exact solutions for Einstein-Maxwell theory with arbitrary dilaton
coupling, and also solutions to the $U(1)^4$ theories that arise from
compactified string/M-theory. Despite the fact that the solutions are
very complicated, physical properties of these black holes, such as
their area, charge, and interaction energy, admit simple expressions.
We also succeed in providing a microscopic description of the
entropy of these black holes using the `effective string' model, and
taking into account the interaction between the
effective string and anti-string.

\end{abstract}

\setcounter{footnote}{0}

\newpage

\section{Introduction}

In spite of the enormous recent progress in understanding the
microscopic origin of the Bekenstein-Hawking entropy using string theory
\cite{SV}, the criticism is sometimes raised that only a very restricted
class of systems are amenable to such a study. Indeed, most of the work
in this area has been performed in situations where either supersymmetry
is present, or at least the deviation from it is, in some sense, small.
In this respect, the exact microscopic calculation of the entropy of the
Schwarzschild black hole remains as the most outstanding problem. It
would also be desirable to have an understanding of more complicated
black hole configurations, and of how the microstates of the black hole
react when submitted to external influences.

As a matter of fact, the study of string and brane physics in
situations far from a supersymmetric state has become a very active
area of research \cite{Sen}. Non-BPS branes and the closely related
brane-antibrane configurations have provided considerable insight into
non-perturbative aspects of string/M-theory, such as dualities and
tachyon condensation. Nevertheless, it is probably fair to say that the
extension to include self-gravity (closed string) effects into the
study of such systems lags far behind their open string description.
Some steps in this direction have been taken in
\cite{senD6,dihole,youm,cet,yol,BOM,Bertolini,bain,Liang}.

In this paper we perform a study of a certain class of black hole
configurations (which can be lifted to configurations of black branes),
whose features provide a connection to several of the above mentioned
issues. Specifically, the systems we study consist of a pair of black
holes at a finite distance, with equal masses, and charges of the same
magnitude but opposite sign. Since these configurations carry an
electric (or magnetic) dipole moment, they are referred to as
\textit{diholes}. In Ref.\ \cite{dihole}, building on the earlier work
of \cite{bonnor,gp,dg,dggh,senD6}, a comprehensive study of such
configurations was carried out for the case when the black holes are
extremal, \ie the horizons are degenerate and the charge is fixed by the
mass. In the present paper, we extend the analysis to the case where the
black holes are not extremal, so the horizons are not degenerate, and
the charge and mass of each black hole are completely independent
parameters. We provide both a macroscopic description, \ie a study of
the solution in classical general relativity, and a string/M-theory
microscopic analysis of the entropy of the configuration, that takes
into account the interaction between the two black holes. In each of
these two veins, we build on previous work and considerably extend it
in several directions.

For the macroscopic description, our starting point is a class of
solutions to Einstein-Maxwell theory, which was built in \cite{Manko}
using a method devised earlier in \cite{sibg}, and which describes a
pair of black holes with arbitrary mass, charge, and rotation
parameters. This very general solution, though, is exceedingly
complicated, and is given in terms of algebraically implicit functions.
Extracting particular cases from the general solution involves a
considerable amount of work, and the analysis of their properties is far
from obvious. For this reason, we have restricted ourselves in this
paper to the subset of parameters that yield the non-extremal diholes
mentioned above. We will be able to find simple expressions for the
area, charge, interaction energy, and other physical properties of the
configurations. Then, we will extend the solutions to theories with
arbitrary dilaton coupling, and also to the theories with four abelian
gauge fields (and three massless scalars) that arise in the low energy
limit of a variety of compactifications of string/M-theory down to four
dimensions. 

With these solutions in hand, we will turn to the microscopic
description of the dihole configuration. Here we follow Ref.\
\cite{mima}, where the entropy of a pair of equally charged,
near-extremal black holes, close to the extremal Majumdar-Papapetrou
solution, is studied. In our case, we will be further away from a
supersymmetric system, since supersymmetry is absent even in the limit
where the black holes are extremal. In fact, the gauge and gravitational
forces between the black holes never cancel each other, instead
they add together. Therefore, even the extremal state reflects in a
non-trivial way the existence of an interaction.

Some comments are in order regarding the connection between the
non-extremal diholes that are the subject of this paper, and the
extremal ones of \cite{bonnor,dg,dihole}. The metric for the former is
much more complicated than the latter. As could be expected, the
extremal dihole is contained in the solutions we study as the limit
where the black hole horizons become degenerate. This means that the
Bonnor solution \cite{bonnor} is a particular case of the solutions in
\cite{Manko}, a point we have explicitly checked. In particular, this
should clear any lingering doubt about whether the dihole interpretation
of Bonnor's solution, as given in \cite{dihole} (as opposed to the
original interpretation as a singular dipole), is the correct one. As we
will see, the extremal limit is a rather subtle one, and this may be one
of the reasons why it took so long to correctly identify the nature of
these solutions. The same remark applies to the interpretation of the
dilatonic dihole solutions constructed in \cite{dg} (which are not a
black hole-white hole pair, as initially suggested). Finally, it is also
worth noting that the new dilatonic solutions we present include, in
particular, the non-extremal extension of the Kaluza-Klein
monopole-antimonopole solution of \cite{gp,dggh}.

\section{Black holes in Weyl coordinates}\label{prewash}

Axisymmetric solutions, such as the ones we are going to study, are
adequately described using Weyl's canonical coordinates. It will be
useful to review the description of several known solutions which are
particular limiting cases of the black diholes that are the subject of
this paper.

For static, electric solutions, a Weyl metric is of the form
\begin{eqnarray}
{\rm d}s^2&=&-f{\rm d}t^2+
f^{-1}[e^{2\gamma}({\rm d}\rho^2+{\rm d}z^2)+\rho^2
{\rm d}\varphi^2]\,,\cr
A_\mu&=&(A_0,0,0,0)\,.
\labell{weyls}
\end{eqnarray}
The functions $f$, $\gamma$ and $A_0$ only depend on $\rho$ and $z$. We
are considering here a solution with an electric potential, but one can
construct a magnetic solution by taking $A_0$ to be the dual magnetic
potential.

\begin{figure}[th]
\centerline{\epsfig{file=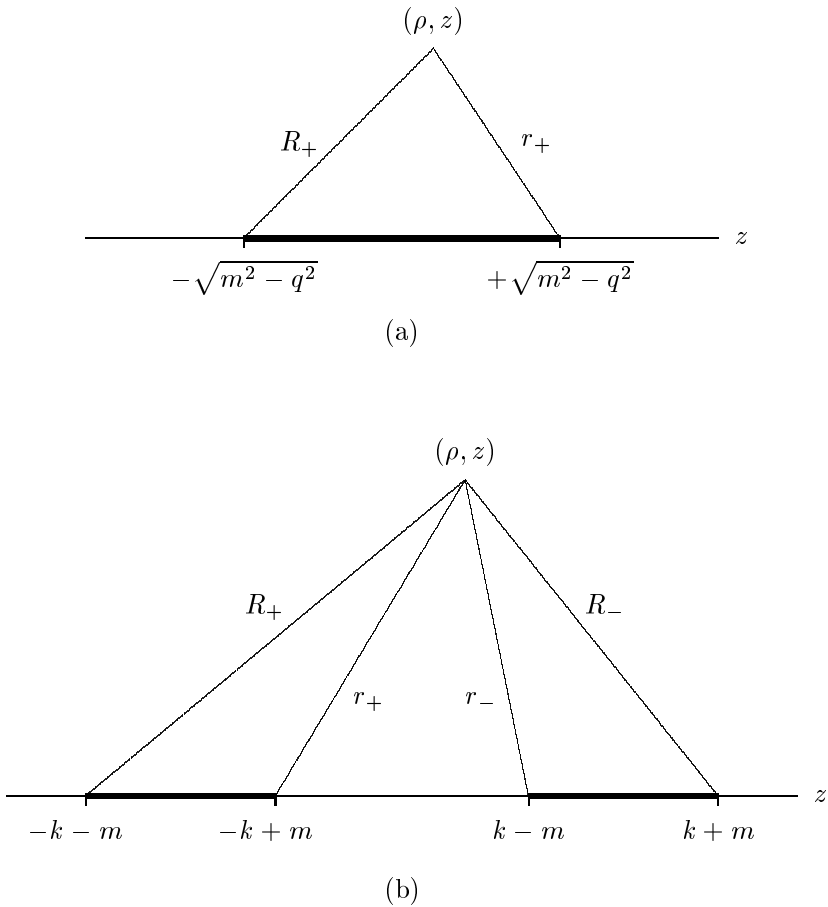,width=11cm}}
\vskip 5mm
\caption{\leftskip1cm\rightskip1cm 
Structure of the axis, and interpretation of $R_\pm$, $r_\pm$,
for the Weyl solutions corresponding to (a) Reissner-Nordstr\"om, (b)
Israel-Khan neutral two-black hole solution.
The black hole horizons correspond to the `rods' marked
by the bold lines.}
\label{examples}
\end{figure}

To start with, let us take the Reissner-Nordstr\"om solution, with 
the familiar metric
\beq
{\rm d}s^2=-{{r^2-2mr+q^2}\over{r^2}}\,{\rm d}t^2+{{r^2{\rm
d}r^2}\over{r^2-2mr+q^2}}+r^2({\rm d}\theta^2+\sin^2\theta\,{\rm
d}\varphi^2)\,,
\labell{RNsol}\end{equation}
and potential $A_0=-q/r$. We transform the $(r,\theta)$ coordinates
into $(\rho, z)$ through
\beqa
\rho&=&\sqrt{r^2-2mr+q^2}\,,\nonumber\\
z&=&(r-m)\cos\theta\,.
\eeqa
Then \reef{RNsol} becomes of the form \reef{weyls} with
\beqa
f&=&{(R_++r_+)^2-4(m^2-q^2)\over (R_++r_++2m)^2}\,,\nonumber\\
e^{2\gamma}&=&{(R_++r_+)^2-4(m^2-q^2)\over 4 R_+r_+}\,,\\
A_0&=&-{2 q\over R_++r_++2m}\,,\nonumber
\eeqa
where we have defined
\beqa
R_+&=&\sqrt{\rho^2+\Big(z+\sqrt{m^2-q^2}\Big)^2}\,,\nonumber\\
r_+&=&\sqrt{\rho^2+\Big(z-\sqrt{m^2-q^2}\Big)^2}\labell{RR}
\eeqa
(the notation $R_+$, $r_+$ has been so chosen for consistency with
future definitions). Figure~\ref{examples}(a) plots the axis of the
solution and the interpretation of the variables $R_+,r_+$. In Weyl's
coordinates, the black holes correspond to `rods' along the axis. For
the Schwarzschild black hole the rod's length is equal to $2m$, but as
charge is added the
length of the rod shrinks. In the extremal limit $m=q$ the rod contracts
to a point, a
fact that makes this limit a rather singular one to take in
Weyl's coordinates.

\bigskip

Now let us consider the Israel-Khan solution \cite{ik}, which describes
a set of neutral black holes along the $z$ axis. For the case of two
black holes of equal mass $m$, with the rods centered at $z=\pm k$, the
solution is
\beqa
f&=&\left({R_++r_+-2m\over R_++r_++2m}\right)
\left({R_-+r_--2m\over R_-+r_-+2m}\right)\,,\nonumber\\
e^{2\gamma}&=&\left({\rho^2+(z+k+m)(z+k-m)+R_+ r_+\over 2 R_+ r_+}\right)
\left({\rho^2+(z-k-m)(z-k+m)+R_- r_-\over 2 R_- r_-}\right)\nonumber\\
&\times&\left({\rho^2+z^2-(k-m)^2+r_+ r_-\over \rho^2
+(z-k-m)(z+k-m)+r_+R_-}\right)
\left({\rho^2+z^2-(k+m)^2+R_+ R_-\over \rho^2
+(z+k+m)(z-k+m)+R_+r_-}\right)\,,\cr
&&\labell{IK}
\eeqa
where now (see Figure~\ref{examples}(b))
\beqa
R_\pm&=&\sqrt{\rho^2+(z\pm (k+m))^2}\,,\nonumber\\
r_\pm&=&\sqrt{\rho^2+(z\pm (k-m))^2}\,.
\eeqa
In this case we have two rods along the axis, each one of length $2m$,
and centers separated by a (coordinate) distance $2k$. The function $\log f$
is indeed the linear superposition of the Newtonian potentials created
by two rods of constant mass per unit length. On the other hand, the function $\gamma$
accounts for the general-relativistic interaction between the black
holes. The black holes are of course expected to attract each other,
and this is encoded in $\gamma$ as follows. Suppose that, in a region
away from the rods, a non-zero value for
\beq
\lim_{\rho\to0}\gamma=\gamma_0\,
\eeq
is obtained as one approaches the axis $\rho=0$. 
Then, with the standard choice for the periodicity of
$\varphi$, \ie $\varphi\sim \varphi+2\pi$, there is a conical
deficit along the axis equal to
\beq
\delta=2\pi(1-e^{-\gamma_0})\,.
\eeq
For the particular metric \reef{IK}, one finds that, outside the rods,
$\gamma_0=0$, hence there is no conical defect there. Instead, on the
segment inbetween the rods the conical angle is
\beq
\delta=-2\pi{m^2\over k^2-m^2}\,,
\eeq
which, being negative, indicates the presence of pressure along the axis
(a `strut') that keeps the black holes apart. 
We are considering the case $k>m$, which corresponds to two non-overlapping rods.
If $k=m$ the solution actually describes a single Schwarzschild black
hole of mass $2m$, whereas if $k<m$ one obtains again two black holes,
but now the roles of $k$ and $m$ are exchanged.
Also, the solution for an individual Schwarzschild black hole can be 
recovered from \reef{IK} by performing the shift $z\to z-k$ and then
taking the limit $k\to\infty$. In this limit, one of the black holes is
pushed away to infinity.

\begin{figure}[th]
\centerline{\epsfig{file=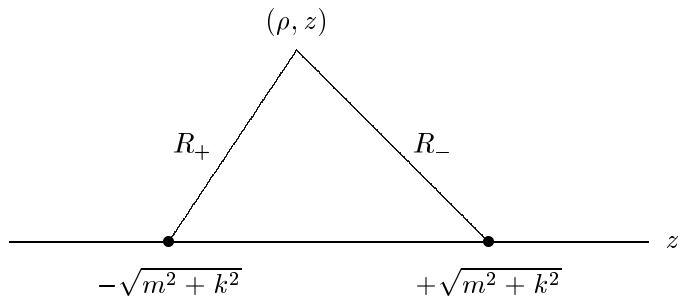,width=9cm}}
\vskip 5mm
\caption{\leftskip1cm\rightskip1cm 
The extremal dihole solution in Weyl coordinates. The extremal
black hole horizons correspond now to the black dots.}
\label{rnweyl}
\end{figure}

Finally, we consider the solution that describes the extremal dihole,
\ie two extremal charged black holes, of the same mass, and charges
equal in magnitude but of opposite sign:
\beqa
f&=&\left[{(R_++R_-)^2-4m^2-{k^2\over m^2+k^2}(R_+-R_-)^2
\over (R_++R_-+2m)^2-{k^2\over m^2+k^2}(R_+-R_-)^2}\right]^2
\,,\nonumber\\
e^{2\gamma}&=&\left[{(R_++R_-)^2-4m^2-{k^2\over m^2+k^2}(R_+-R_-)^2
\over 4 R_+R_-}\right]^4\,,\labell{extdihole}\\ 
A_0&=&-{4 m k\over \sqrt{m^2+k^2}}{R_+-R_-\over (R_++R_-+2m)^2-{k^2\over
m^2+k^2}(R_+-R_-)^2}\,,\nonumber
\eeqa
with, now,
\beq
R_\pm=\sqrt{\rho^2+(z\pm\sqrt{m^2+k^2})^2}\,,
\eeq 
see Figure~\ref{rnweyl}. In this case the black holes are extremal, \ie
their horizons are degenerate. Hence each black hole is represented by a
point on the axis. For more details about this solution, see
\cite{dihole}. 

\section{The non-extremal dihole}\label{thething}

The class of solutions of Einstein-Maxwell theory that describes two
black holes along the axis, with arbitrary mass, charge, and rotation
parameters, was presented in \cite{Manko}, following a method devised by
Sibgatullin \cite{sibg}. Such solutions are indeed
very complicated, and are given in terms of a number of algebraically
implicit functions---a considerable amount of work is required to
recover particular cases from them. We are interested in the solution
for two non-extremal black holes, of equal mass, and charge of the same
magnitude but opposite sign. Henceforth this solution will be
referred to as a non-extremal dihole. In the notation of \cite{Manko},
our choice of
parameters is
\begin{eqnarray}
m_1=m_2=m,&\quad&q_1=-q_2=q\,,\cr
a_1=a_2=0,&\quad&z_1=-z_2=k\,.
\end{eqnarray}
Then, the solution
in Weyl's canonical form \reef{weyls},
\begin{eqnarray}
{\rm d}s^2&=&-f{\rm d}t^2+
f^{-1}[e^{2\gamma}({\rm d}\rho^2+{\rm d}z^2)+\rho^2
{\rm d}\varphi^2]\,,
\end{eqnarray}
is\footnote{Note that there is a misprint in (3.17) of
\cite{Manko}: $z+m$ should be $z$. This equation is needed to 
show that $\omega=0$, \ie the $g_{t\varphi}$ component of the
metric vanishes.}
\begin{equation}
\label{f}
f={\mathcal{A}^2-\mathcal{B}^2+\mathcal{C}^2\over
(\mathcal{A}+\mathcal{B})^2},\qquad
e^{2\gamma}={\mathcal{A}^2-\mathcal{B}^2+\mathcal{C}^2
\over K_0R_+R_-r_+r_-},\qquad
A_0=-{\mathcal{C}\over \mathcal{A}+\mathcal{B}}\,,
\labell{solution}\end{equation}
where $\mathcal{A}$, $\mathcal{B}$ and $\mathcal{C}$ are given by
\begin{eqnarray}
\mathcal{A}&=& -(R_+ -R_-)(r_+-r_-)[(\kappa_+^2 +
\kappa_-^2)(m^4+\kappa_+^2\kappa_-^2)-4 m^2\kappa_+^2\kappa_-^2]\cr
&&-2(\kappa_+^2-\kappa_-^2)(R_+R_-+r_+r_-)[m^4-(\kappa_+\kappa_-)^2]\cr
&&+(\kappa_+^2-\kappa_-^2)(R_++R_-)(r_++r_-)[m^4+(\kappa_+\kappa_-)^2]\,,
\cr
\mathcal{B}&=&
4m(\kappa_+^2-\kappa_-^2)\kappa_+\kappa_-[(R_++R_-+r_++r_-
)\kappa_+\kappa_-
-(R_++R_--r_+-r_-)m^2]\,,\cr
\mathcal{C}&=&
4kq\kappa_+\kappa_-[(R_+-R_-+r_+-r_-)(\kappa_+^2-m^2)\kappa_--(R_+-R_--
r_++r_-)(\kappa_-^2-m^2)\kappa_+]\,.\cr
&&
\labell{ABC}\end{eqnarray}
Here, we have defined
\begin{equation}
R_\pm\equiv\sqrt{\rho^2+(z\pm(\kappa_++\kappa_-))^2}\,,\qquad
r_\pm\equiv\sqrt{\rho^2+(z\pm(\kappa_+-\kappa_-))^2}\, .
\labell{rpm}
\end{equation}

The solution depends on three parameters---in physical terms, the mass,
charge and separation between the black holes. We have found it
convenient to choose these to be $m$, $\kappa_+$ and $\kappa_-$, all of
them positive. In terms of the parameters $q$ and $k$ introduced in
\cite{Manko}, they are expressed as
\beq
\kappa_\pm={1\over 2}\bigg(\sqrt{m^2+k^2+ 2k\sqrt{m^2-
q^2}}\pm\sqrt{m^2+k^2-
2k\sqrt{m^2-q^2}}\bigg)\, .
\labell{kappapm}\eeq
Conversely,
\beqa
k^2&=&\kappa_+^2+\kappa_-^2-m^2,\nonumber\\
kq&=&\sqrt{(\kappa_+^2-m^2)(m^2-\kappa_-^2)}\, .
\label{converse}
\eeqa

\begin{figure}
\centerline{\epsfig{file=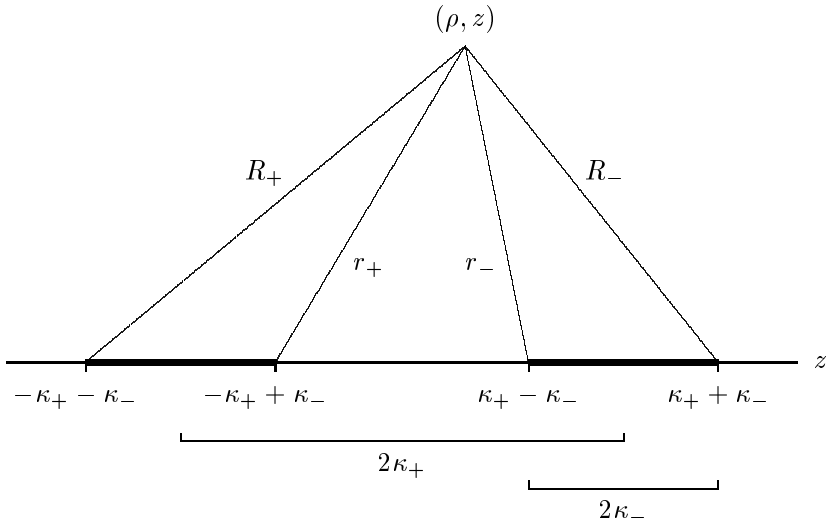,width=11cm}}
\vskip 5mm
\caption{\leftskip1cm\rightskip1cm 
Structure of the axis, and interpretation of $R_\pm$, $r_\pm$,
\reef{rpm},
and $\kappa_\pm$.
The black hole horizons correspond to the `rods' marked by the bold
lines.}
\label{axis}
\end{figure}

The black hole horizons lie on the axis $\rho=0$ at $-\kappa_-\leq
z\pm\kappa_+\leq\kappa_-$, see Figure \ref{axis}. 

We will restrict the parameters to satisfy $\kappa_-\leq m<\kappa_+$.
With this choice one obtains a real solution describing two black holes.
Then, as is clear from the figure, $\kappa_+$ controls the separation
between the black holes, while $\kappa_-$ is a `non-extremality'
parameter. The case $\kappa_+\leq m<\kappa_-$ actually also describes
two black holes, since it can be mapped to the former situation by
exchanging $\kappa_+\leftrightarrow\kappa_-$ and also at the same time
$r_+\leftrightarrow r_-$. These solutions, where $\kappa_+$ and
$\kappa_-$ exchange their roles as giving the separation and length of
the rods, are therefore redundant and we will not need to consider them
separately. One may also ask what happens when the two oppositely
charged black holes touch, $\kappa_+=\kappa_-$. In this case the
solution turns out to be a neutral one, since \reef{kappapm} then
implies $q=0$ (hence $\mathcal{C}=0$), and $\kappa_\pm=m$. The solution
is in fact a Schwarzschild black hole of mass $2m$. Finally, we remark
that neither $q$ nor $k$ are the physical charge or separation between
the black holes, but simply parameters that, in the limit of large
separation between the holes, approximate these two quantities. The
actual physical charge will be determined below, while the proper
distance between the black holes does not appear to be too relevant, and
at any rate will not be needed in the following.

It still remains to determine $K_0$, which is a quantity that can depend
on the parameters of the solution but not on the coordinates. Which
value one chooses for $K_0$ is dictated by the structure of conical
singularities along the axis. As discussed in the previous section, in a
portion of the axis away from the horizons, if the periodicity of
$\varphi$ is $\Delta\varphi$, and $\gamma_0\equiv\gamma|_{\rho\to 0}$,
then there will be a conical deficit 
\beq
\delta=2\pi-\Delta\varphi\; e^{-\gamma_0}\, .
\labell{condef}\eeq
A direct calculation yields, for the portions inbetween the black holes,
and outside them,
\beqa
e^{-\gamma_0}&=&{K_0^{1/2}\over
8\kappa_-^2(\kappa_+^2-m^2)^2},\quad |z|<\kappa_+-\kappa_-\labell{inside}\\
e^{-\gamma_0}&=&{K_0^{1/2}\over
8\kappa_+^2\kappa_-^2(\kappa_+^2-\kappa_-^2)},\quad |z|>\kappa_++\kappa_-
\labell{outside}
\eeqa
If we choose to cancel the defect at infinity then we set 
\beq
K_0^{1/2}=8\kappa_+^2\kappa_-^2(\kappa_+^2-\kappa_-^2)\, ,
\labell{chooseK0}\eeq
together with $\Delta\varphi=2\pi$ (only the product
$\Delta\varphi\;K_0^{1/2}$ is significant). In that case, the
conical angle on the
axis is negative (\ie an excess angle, or `strut'),
\beq
\delta_{\mathrm{strut}}=-2\pi{m^2(\kappa_+^2-m^2)+\kappa_+^2(m^2-\kappa_-^2)
\over(\kappa_+^2-m^2)^2}\,.
\labell{dstrut}\eeq
On the other hand, if, instead, we had chosen to remove the singularity
from the axis inbetween the black holes, we would have found a conical
deficit (a `cosmic string') extending out to infinity,
\beq
\delta_{\mathrm{string}}=2\pi{m^2(\kappa_+^2-m^2)+\kappa_+^2(m^2-\kappa_-
^2)\over\kappa_+^2(\kappa_+^2-\kappa_-^2)}\, .
\labell{dstring}\eeq

In the following we will choose to have a strut singularity inbetween
the black holes, hence $\Delta\varphi=2\pi$, and $K_0$ as in
\reef{chooseK0}.

The solutions described in the previous section can be seen to be
recovered in the following limits:

\begin{itemize}

\item Individual non-extremal Reissner-Nordstr\"om: this corresponds to
infinite separation between the black holes, \ie $k\to\infty$,
$\kappa_+\to k$, with $m$ and $\kappa_-\to\sqrt{m^2-q^2}$ finite. In
order to leave one of the black holes at a finite distance while the other
is sent to infinity, first shift $z\to z-\kappa_+$, and then take
$k\to\infty$.

\item Israel-Khan (two Schwarzschild black holes): $q=0$, $\kappa_+=k$,
$\kappa_-=m$. 

\item Extremal dihole: $q=m$, $\kappa_+=\sqrt{m^2+k^2}$, $\kappa_-=0$.

\end{itemize}

The limit for the extremal dihole is particularly tricky, since
the rods along the axis shrink to points, $\kappa_-\to 0$, and
$R_\pm\to r_\pm$. Since $\mathcal{A}$, $\mathcal{B}$ and $\mathcal{C}$ 
all vanish with leading order 
$O(q-m)^2$, the $q\rightarrow m$ limit must be taken with care.
It turns out that the expressions simplify considerably by changing
to prolate spheroidal coordinates, $(x,y)$, with 
$2x\equiv (R_++R_-)/\sqrt{m^2+k^2}$, 
$2y\equiv (R_+-R_-)/\sqrt{m^2+k^2}$. 
After lengthy manipulations one then recovers the same expressions
as in \reef{extdihole}.

There is a last limit of some interest, namely the limit $k=0$. This
implies $\kappa_+=m$, $\kappa_-=0$, and the charge vanishes. The
solution reduces to the Darmois solution, with 
\begin{equation}
f=\left({R_++R_--2m\over R_++R_-+2m}\right)^2. 
\end{equation} 
For this solution, not only do the rods shrink to naked null
singularities, but also the segment inbetween them develops an infinite
conical defect. It is also the same solution as is obtained in the
limit $k=0$ of the extremal dihole, where its dipole moment vanishes.
The significance of this solution in this context is somewhat obscure.

\section{Physical properties of the dihole}\label{properties}

In spite of its daunting aspect, the physical properties of the solution
\reef{solution} can be computed in a straightforward manner and will
take simple forms in terms of $m$ and $\kappa_\pm$. First, note that by
examining the solution at asymptotically large distances one quite
easily sees that the total mass is $2m$ and the electric dipole moment
$2 k q$. Other properties, such as the area, temperature and charge of
the black holes depend on the form of the solution close to the
horizons, and will be calculated now.

Let us start with the computation of the area of the black holes. For
definiteness, one can focus on the black hole that lies along the
segment $-\kappa_-\leq z-\kappa_+\leq\kappa_-$, the area of the other
one being obviously the same. This area is given by
\beq
A_\mathrm{bh}=\int
{\rm d}\varphi\;{\rm d}z\sqrt{g_{zz}g_{\varphi\varphi}}|_{\rho=0}=4\pi\kappa_-
\left({\rho e^\gamma\over f}\right)_{\rho=0}\, .
\labell{areagen}\eeq
The limit $\rho\to 0$ has to be taken with a bit of care, but does not
present much complication. The calculations are long but
straightforward,
and one readily finds
the pleasantly simple result
\beq
\left({\rho e^\gamma\over f}\right)_{\rho=0}={(\kappa_++m)^2(\kappa_-
+m)^2\over \kappa_+\kappa_-(\kappa_++\kappa_-)}\, .
\eeq
Hence the area of each black hole is 
\beq
A_\mathrm{bh}={4\pi\over\kappa_+}{(\kappa_++m)^2(\kappa_-
+m)^2\over \kappa_++\kappa_-}\, .
\labell{area}\eeq

As a check, the areas for the three limiting cases above (individual
Reissner-Nordstr\"om black hole, Israel-Khan two Schwarzschild black holes,
and extremal dihole) can be readily recovered from here.

The temperature $T=\beta^{-1}$ of the black holes can be calculated, by
continuing to Euclidean time, to be
\beqa
\beta&=&4\pi\left({e^\gamma\over\partial_\rho f}\right)_{\rho=0}\nonumber\\
&=&2\pi
{(\kappa_++m)^2(\kappa_-+m)^2\over\kappa_+\kappa_-(\kappa_++\kappa_-
)}\,.
\labell{temp}\eeqa
In the above limits it reproduces again the expected results.
Note that one obtains the simple
relation
\beq
A_\mathrm{bh}=2\beta\kappa_-\,,
\labell{areatemp}\eeq
which indeed is a straightforward consequence of \reef{areagen} and
\reef{temp}.

Next we will compute the charge of each black hole. To do so, we employ
Gauss' law $Q={1\over 4\pi}\int_{S^2}*F$, taking the (topological)
sphere $S^2$ to
be a surface at constant small $\rho$ closely surrounding the black hole
horizon. In that case, it is straightforward to find that the black hole
charge is given by
\beqa
Q&=&{1\over 4\pi}\int{\rm d}\varphi\int_{\kappa_+-\kappa_-}^{
\kappa_++\kappa_-}{\rm d}z\left({\rho\over f}\partial_\rho
A_0\right)_{\rho=0}\nonumber\\
&=&{\sqrt{(\kappa_+^2-m^2)(
m^2-\kappa_-^2)}\over \kappa_+-m}\, .
\eeqa
Notice that, with \reef{converse}, this is simply
\beq
Q={kq\over \kappa_+-m}\, .
\labell{charge}\eeq
Later we will also require the value of the electric potential at the
horizons, $\Phi$. In order to fix the arbitrary additive constant in the
potential $A_0$, in \reef{solution} we have made the choice that $A_0$
vanishes at asymptotic infinity. This is a reasonable choice, since in
contrast to the situation for an individual black hole, or for a number of
equally charged black holes, one cannot choose the arbitrary constant in
the potential to make the potentials vanish simultaneously on both
horizons. Then, from \reef{solution}, it follows that
\beqa
\Phi&=&\sqrt{(\kappa_+-m)(m-\kappa_-)\over(\kappa_++m)(m+\kappa_-
)}\nonumber\\
&=&{m-\kappa_-\over Q}\, .
\labell{potential}\eeqa
$Q$ and $\Phi$ each change sign from one black hole to the other, but
the product $Q\Phi$ has the same sign for both.

Finally, we calculate the interaction energy between the black holes.
A straightforward hamiltonian
analysis shows that the energy
associated to a conical defect $\delta$ along the $z$ axis is
\beq
E_\delta=-{\delta\over 4}\int{\rm d}z\; N\sqrt{g_{zz}}|_{\rho=0}\,,
\eeq
where $N=\sqrt{-g_{tt}}$ is the lapse function for the static metric
under consideration. Identifying this energy with the interaction energy
between the black holes, $V_\mathrm{int}\equiv E_\delta$, we get
\beqa
V_\mathrm{int}&=&{\kappa_+-\kappa_-\over 2}(e^{\gamma_0}-1)\nonumber\\
&=&-{\kappa_+^2(\kappa_+^2-\kappa_-^2)-(\kappa_+^2-m^2)^2\over
2\kappa_+^2(\kappa_++\kappa_-)}\, .
\labell{vint}\eeqa
The value of $\gamma_0$ used here is the one
obtained from \reef{inside} and \reef{chooseK0}. At large separation, \ie
large $\kappa_+$, the interaction energy becomes
\beq
V_\mathrm{int}\simeq -{m^2+q^2\over 2\kappa_+}\,,
\eeq
which is indeed the result expected for the gravitational and electric
interaction of two point masses and charges at a distance $\sim
2\kappa_+$.

The remarkable conclusion of this section is that, even if the metric
for the non-extremal dihole is a very complicated one, we have managed
to obtain simple expressions for all the physical properties of the
configuration.

\section{Dilatonic and string/M-theory diholes}\label{stringy}

\subsection{Dilatonic dihole solution}

We would now like to generalise the non-extremal dihole solution 
to a solution of Einstein-Maxwell-dilaton theory: 
\beq
I={1\over 16\pi G}\int{\rm d}^4x\sqrt{-g}\left(R-2(\nabla\phi)^2-
e^{-2\alpha\phi}F^2\right),
\eeq
with arbitrary dilaton coupling $\alpha$. This is necessary should we 
want to make contact with Kaluza-Klein theory ($\alpha=\sqrt{3}$) or a
particular low-energy effective limit of string theory (the simplest,
$\alpha=1$). 

It was shown in \cite{Liang} how one can generate a static, 
axisymmetric solution of the Einstein-Maxwell-dilaton equations 
starting from a stationary, axisymmetric solution of the vacuum 
Einstein equations. Indeed, this procedure was used in \cite{dg} to 
generate the extreme dilatonic dihole solution (for general $\alpha$) 
from the Kerr solution. It is straightforward to modify this procedure 
to generate a general-$\alpha$ solution starting from the corresponding
$\alpha=0$ (pure Einstein-Maxwell) solution, without any reference to
the seed vacuum Einstein solution. We shall now outline this
procedure, leaving the reader to fill in the details using the results
of \cite{Liang}.

Suppose we have an electrically charged\footnote{The same results would
apply to a magnetically charged solution, provided we replace $A_0$ by
the magnetic potential $A_3$ and set $\phi\rightarrow-\phi$. However,
this solution-generating procedure fails for dyonic solutions.} 
solution of Einstein-Maxwell theory specified by $f$, $\gamma$ and 
$A_0$, as in \reef{weyls}. Then the general dilatonic solution has 
a new $f'$, $\gamma'$, $A_0'$, and dilaton $\phi$, given by
\beqa
f'&=&f^{1\over1+\alpha^2}e^{-2\alpha\tilde\phi},\cr
\gamma'&=&{1\over1+\alpha^2}\,\gamma+\tilde\gamma\,,\cr
A_0'&=&{1\over\sqrt{1+\alpha^2}}\,A_0\,,\cr
e^{2\phi}&=&f^{\alpha\over1+\alpha^2}e^{2\tilde\phi}\,,
\labell{solgen}
\eeqa
where $\tilde\phi$ is a harmonic function satisfying 
\beq
{\partial^2\tilde{\phi}\over\partial\rho^2}+
{\partial^2\tilde{\phi}\over\partial z^2}+{1\over\rho}
{\partial\tilde{\phi}\over\partial\rho}=0\,,
\eeq
and $\tilde\gamma$ is given in terms of $\tilde\phi$ by
\beqa
{\partial\tilde{\gamma}\over\partial\rho}&=&
(1+\alpha^2)\rho\left[\bigg({\partial
\tilde{\phi}\over\partial\rho}\bigg)^2-
\bigg({\partial\tilde{\phi}\over\partial z}\bigg)^2\right]\,,\cr
{\partial\tilde{\gamma}\over\partial z}&=&
2(1+\alpha^2)\rho\,{\partial\tilde{\phi}\over\partial\rho}
{\partial\tilde{\phi}\over\partial z}\,.
\eeqa

This solution-generating procedure can be used, for instance, to obtain
the extreme dilatonic dihole solution starting from the Bonnor solution.
It turns out that $\tilde\phi=0$ in this case \cite{Liang}. However,
this is not necessarily so in general. For example, we must take
\beq
e^{-2\tilde\phi}=
\left({R_++r_+-2\sqrt{m^2-q^2}\over R_++r_+
+2\sqrt{m^2-q^2}}\right)^{\alpha\over1+\alpha^2},
\eeq
if we want to generate the non-extreme dilatonic black hole solution
from the Reissner-Nordstr\"om solution. In this case, we obtain (in
Weyl  coordinates)
\beqa
f&=&{R_++r_+-2\sqrt{m^2-q^2}\over R_++r_++2m}\left({R_++r_++2\sqrt{m^2-q^2}
\over R_++r_++2m}\right)^{1-\alpha^2\over1+\alpha^2},\cr
e^{2\gamma}&=&{(R_++r_+)^2-4(m^2-q^2)\over 4 R_+r_+}\,,\cr
A_0&=&-{1\over\sqrt{1+\alpha^2}}\,{2 q\over R_++r_++2m}\,,\cr
e^{2\phi}&=&\left({R_++r_++2\sqrt{m^2-q^2}
\over R_++r_++2m}\right)^{2\alpha\over1+\alpha^2}\,,
\labell{dilatonbh}
\eeqa
which indeed corresponds to the dilatonic black hole solution
\cite{dilbh}.

Applying the solution-generating procedure \reef{solgen} to the 
non-extremal dihole solution \reef{solution}, we obtain
\beqa
{\rm d}s^2&=&-f{\rm d}t^2+f^{-1}[e^{2\gamma}({\rm d}\rho^2+{\rm d}z^2)
+\rho^2{\rm d}\varphi^2]\,,\cr
A_0&=&-{1\over\sqrt{1+\alpha^2}}\,{{\cal C}\over{\cal A}+{\cal B}}\,,\cr
e^{2\phi}&=&\left[{{\cal A}^2-{\cal B}^2+{\cal C}^2\over({\cal A}+
{\cal B})^2}\right]^{\alpha\over1+\alpha^2}e^{2\tilde\phi}\,,
\labell{dilatondihole}
\eeqa
where
\beq
f=\left[{{\cal A}^2-{\cal B}^2+{\cal C}^2\over({\cal A}+
{\cal B})^2}\right]^{1\over1+\alpha^2}
e^{-2\alpha\tilde\phi},\qquad
e^{2\gamma}=\left[{{\cal A}^2-{\cal B}^2+{\cal C}^2\over 
K_0R_+R_-r_+r_-}\right]^{1\over1+\alpha^2}e^{2\tilde\gamma},
\eeq
and $\mathcal{A}$, $\mathcal{B}$, $\mathcal{C}$ are the same as in
\reef{ABC}. We shall choose the harmonic function $\tilde\phi$ to be
given by
\beq
e^{-2\tilde\phi}=
\left[\left({R_++r_+-2\kappa_-\over R_++r_++2\kappa_-}\right)
\left({R_-+r_--2\kappa_-\over R_-+r_-+2\kappa_-}\right)
\right]^{\alpha\over1+\alpha^2},
\labell{tildephi}
\eeq
from which we derive
\beqa
e^{2\tilde\gamma}&=&\Bigg[{\rho^2+(z+\kappa_++\kappa_-)
(z+\kappa_+-\kappa_-)+R_+r_+\over2R_+r_+}\cr
&&\times\,{\rho^2+(z-\kappa_+-\kappa_-)(z-\kappa_++\kappa_-)+R_-r_-
\over2R_-r_-}\cr
&&\times\,{\rho^2+z^2-(\kappa_++\kappa_-)^2+R_+R_-\over
\rho^2+(z+\kappa_++\kappa_-)(z-\kappa_++\kappa_-)+R_+r_-}\cr
&&\times\,{\rho^2+z^2-(\kappa_+-\kappa_-)^2+r_+r_-\over
\rho^2+(z+\kappa_+-\kappa_-)(z-\kappa_+-\kappa_-)+r_+R_-}
\Bigg]^{\hbox{$\alpha^2\over1+\alpha^2$}}\,,
\eeqa
using the results of \cite{ik}.
With this, we have the complete non-extremal dilatonic dihole solution. 
It correctly reduces to the solution of \cite{dg} in the extreme limit.
Furthermore, the individual non-extremal dilatonic black holes 
(given by \reef{dilatonbh}) can be
recovered in the infinite separation limit, thus confirming that 
the choice of $\tilde\phi$ in \reef{tildephi} is the correct one.

We briefly turn to some physical properties of this solution.
Now, we have $e^{\tilde\gamma}|_{\rho=0}=1$ on the axis outside the
black holes. To cancel the conical defect at infinity, $K_0$ therefore 
has to assume the same value as in the Einstein-Maxwell case,
\reef{chooseK0}. 
With this value, the area of each black hole is
\beq
A_\mathrm{bh}={4\pi\over\kappa_+}\,
{(4\kappa_-^2)^{\alpha^2\over1+\alpha^2}
[(\kappa_++m)(\kappa_-+m)]^{2\over1+\alpha^2}
\over(\kappa_++\kappa_-)^{1-\alpha^2\over1+\alpha^2}}\,.
\eeq
Note that when $\alpha\neq0$, the area vanishes in the extreme limit as
expected. It also gives the correct expression in the limit of infinite
separation. We can also read off from $A_\mathrm{bh}$, the inverse
temperature $\beta$ of each black hole using \reef{areatemp}.

The electric charge of each black hole is 
\begin{eqnarray}
Q&=&{1\over 4\pi}\int {\rm d}\varphi\int_{\kappa_+-\kappa_-}^{\kappa_+
+\kappa_-}{\rm d}z\left({\rho\over f}e^{2\alpha\phi}\partial_\rho
A_0\right)_{\rho=0}\nonumber\\
&=&{1\over\sqrt{1+\alpha^2}}\,{\sqrt{(\kappa_+^2-m^2)(m^2-\kappa_-
^2)}\over \kappa_+-m}\,,
\end{eqnarray}
while the value of the potential at the horizon is 
\beq
\Phi=\sqrt{{1\over1+\alpha^2}\,{(\kappa_+-m)(m-\kappa_-)\over(\kappa_++m)
(m+\kappa_-)}}\,.
\eeq
Lastly, we note that $e^{\tilde\gamma}|_{\rho=0}=\Big[{\kappa_+^2-
\kappa_-^2\over\kappa_+^2}\Big]^{\alpha^2\over1+\alpha^2}$ on the axis 
inbetween the black holes. The interaction energy between the black holes 
can then be calculated to be
\beq
V_\mathrm{int}=-{\kappa_+^2(\kappa_+^2-\kappa_-^2)-(\kappa_+^2-
m^2)^{2\over1+\alpha^2}(\kappa_+^2-\kappa_-^2)^{2\alpha^2\over1+\alpha^2}
\over 2\kappa_+^2(\kappa_++\kappa_-)}\, .
\eeq

To find the corresponding magnetically charged dihole solution, 
we have to perform the duality transformation 
$F\rightarrow\ast F$, $\phi\rightarrow-\phi$, of which the former 
can be written as
\begin{equation}
{\partial A_3\over\partial\rho}=\rho f^{-1}e^{-2\alpha\phi}\,
{\partial A_0\over\partial z}\,,
\qquad
{\partial A_3\over\partial z}=-\rho f^{-1}e^{-2\alpha\phi}\,
{\partial A_0\over\partial\rho}\,.
\labell{duality}
\end{equation}
Unfortunately, integrating these expressions to obtain the magnetic
potential $A_3$ appears to be a formidable task, and we have not been 
able to obtain an explicit expression for it. We end off with the remark
that the magnetic dihole solution in Kaluza-Klein theory 
generalises the monopole-antimonopole solution found in \cite{gp,dggh}.
It can also be uplifted to Type IIA string theory, to describe
a non-extremal (black) D6-anti-D6-brane configuration, thereby extending the 
extremal solution in \cite{dggh,senD6}.

\subsection{$U(1)^4$ dihole solution}

In this subsection, we turn to a theory consisting of four abelian 
gauge fields and three scalar fields, given by the action
\beqa
\label{u14action}
I&=&{1\over 16\pi G}\int{\rm d}^4x\sqrt{-g}\,\biggl\{R-{1\over 2}\left[
(\partial\eta)^2 + (\partial\sigma)^2 +(\partial\tau)^2\right]\nonumber
\\
&&\qquad\qquad-{e^{-\eta}\over 4}\left[ e^{-\sigma-\tau}F_{(1)}^2
+ e^{-\sigma+\tau}F_{(2)}^2
+e^{\sigma+\tau}F_{(3)}^2 +e^{\sigma-\tau}F_{(4)}^2\right]\biggr\}\,.
\eeqa
This so-called $U(1)^4$ theory is of wide interest because it emerges as a
consistent truncation of a number of different compactifications of
low-energy string theory and M-theory \cite{fourcharge}. The one that we
shall focus on below consists of three 5-branes in M-theory intersecting
on a common string, along which momentum is flowing \cite{intbrn}. 

Solutions to this theory describing a pair of non-extremal black holes
accelerating apart were constructed in \cite{compo}. Now we aim to find
the non-extremal dihole solution in this theory. The individual black
holes must carry equal and opposite values of four charges under each of
the gauge fields, two of which are electric and the other two magnetic.
Hence the dihole solution will depend on a total of six parameters: four
for each of the charges, one for the total mass, and one for the
distance between the black holes. It will be convenient to choose these
parameters to be $\kappa_\pm$, and $m_i$, $i=1,\dots,4$. Notice that
$\kappa_\pm$ do not depend on the index $i$ labeling each $U(1)$. 

Bearing the latter fact in mind, it is possible to find a simple factorised 
form for the dihole line element:
\begin{equation}
{\rm d}s^2=-(f_1f_2f_3f_4)^{1\over4}{\rm d}t^2+(f_1f_2f_3f_4)^{-{1\over4}}
\left[e^{{1\over2}(\gamma_1+\gamma_2+\gamma_3+\gamma_4)} 
({\rm d}\rho^2+{\rm d}z^2)
+\rho^2{\rm d}\varphi^2\right]\,,
\labell{u14metric}
\end{equation}
where                                     
\beq
f_i\equiv {{\cal A}_i^2-{\cal B}_i^2+{\cal C}_i^2\over({\cal A}_i+
{\cal B}_i)^2},
\qquad e^{2\gamma_i}\equiv{{\cal A}_i^2-
{\cal B}_i^2+{\cal C}_i^2\over K_0 R_+R_-r_+r_-}.
\labell{tsigma}
\eeq
${\cal A}_i$, ${\cal B}_i$ and ${\cal C}_i$ are defined as in \reef{ABC}, 
but with $m$ replaced by $m_i$. Again, $K_0$ should be chosen as in
\reef{chooseK0} if one requires regularity on the axis outside the
dihole (notice this is consistent with $K_0$ not having an index $i$).
The three scalar fields are given by
\begin{equation}
e^{-2\eta}=\left({f_1f_3\over f_2f_4}\right)^{1\over2}\,,\qquad
e^{-2\sigma}=\left({f_1f_4\over f_2f_3}\right)^{1\over2}\,,\qquad
e^{-2\tau}=\left({f_1f_2\over f_3f_4}\right)^{1\over2}\,.
\end{equation}
Furthermore, the electric potentials $A_{(2)}$ and $A_{(4)}$ are
\begin{equation}
A_{(i)}=-{{\cal C}_i\over{\cal A}_i+{\cal B}_i}\,,\qquad i=2,4\,,
\labell{ai}\end{equation}
while the magnetic potentials $A_{(1)}$ and $A_{(3)}$ are given 
implicitly by applying the analogue of the duality transformation 
(\ref{duality}) on the corresponding electric potentials. We will not
need their explicit forms.

This solution has been checked to satisfy the field equations coming from
\reef{u14action}. Another consistency check involves setting 1, 2, 3 or 4 
of the charges to be non-zero and equal. As is well known, this should 
reduce to the dilatonic dihole solution \reef{dilatondihole} for 
$\alpha=\sqrt3$, 1, $1\over\sqrt{3}$ or 0, respectively \cite{fourcharge}. 
Indeed, this can readily be checked if we recall that in the chargeless 
limit $\kappa_-=m_i$,
\beq
f_i=\left({R_++r_+-2\kappa_-\over R_++r_++2\kappa_-}\right)
\left({R_-+r_--2\kappa_-\over R_-+r_-+2\kappa_-}\right).
\eeq
This expression is of the right form to reproduce the 
$e^{-2\alpha\tilde\phi}$ terms in the metric for each of the four cases. 
A similar analysis of the $\Sigma_i$ terms shows that they 
correctly give the $e^{2\tilde\gamma}$ term in the metric as well. 

The total mass of the solution, as measured at infinity, is ${1\over
2}\sum_{i=1}^4m_i$, and the dipole moment for each of the gauge fields
is $2k q_i=2\sqrt{(\kappa_+^2-m_i^2)(m_i^2-\kappa_-^2)}$. Finally, the
area, charge and potential at the horizon for each of the two black
holes, and the interaction energy between them, are
\beqa
A_\mathrm{bh}&=&4\pi{\prod_{i=1}^4\sqrt{(\kappa_++m_i)(\kappa_-+m_i)}\over
\kappa_+(\kappa_++\kappa_-)}=2 \beta\kappa_-\, ,\
\labell{composite}\\
Q_i&=&{\sqrt{(\kappa_+^2-m_i^2)(m_i^2-\kappa_-^2)}\over
\kappa_+-m_i}\, ,\\ 
\Phi_i&=&{m_i-\kappa_-\over Q_i}\, ,\\
V_\mathrm{int}&=&-{\kappa_+^2(\kappa_+^2-\kappa_-^2)-\prod_{i=
1}^4\sqrt{\kappa_+^2-m_i^2}\over
2\kappa_+^2(\kappa_++\kappa_-)}\, .
\eeqa

\section{Statistical mechanics of the dihole}\label{thermo}

We now want to find a consistent microscopic description, within
string/M-theory, of the entropy of the dihole in the case where the two
oppositely charged black holes are extremal or nearly extremal. 

Our study will be along the lines of the beautiful analysis in
\cite{mima}, where a configuration of two black holes with equal charges
of \textit{the same sign} is studied. There is a significant difference
between our dihole configuration and the system in \cite{mima}. Namely,
in the case that the black holes have charges of the same sign, the
force between them vanishes in the extremal limit. Therefore, in that
case the near-extremal limit is also one where the interaction between
the black holes is small. The ground state of this system is a BPS
state. This is not the case for the configuration we study: the force
between both black holes does not cancel in the extremal limit, and all
the supersymmetries are broken. Thus, we will need to study the extremal
dihole first. 

We will work in the approximation where the distance between the black
holes is large,
\beq
\kappa_+\gg m,\kappa_-\, .
\labell{longdist}\eeq
In this case, the interaction between the black holes is weak. It is
unclear whether the solution will be reliable, for thermodynamic and
semiclassical analysis, when the interaction energy that is stored in
the strut is of the order of, or larger than, the mass of the black
holes. As in \cite{mima}, we regard the strut as a sort of boundary
condition on the system that accounts for the interaction. If the strut
singularity becomes too strong, the distortion introduced may invalidate
the approach. Recall that, indeed, in the limit $k=0$ the horizons
disappear and are replaced by a naked singularity of diverging
curvature.

\subsection{Energy and entropy of the extremal dihole}

As a preliminary for the microscopic analysis of the near-extremal
dihole, we will discuss here the energetics of the configuration when
the black holes are extremal, \ie their horizons are degenerate. For
simplicity, we will consider the non-dilatonic case. Recall that this
configuration is described by Bonnor's solution, which can be obtained
from \reef{solution} as the particular case where
\beq
q=m,\quad \kappa_+=\sqrt{m^2+k^2},\quad \kappa_-=0\, .
\eeq
Hence, the area and charge of each black hole become
\beq
A_\mathrm{bh}=4\pi m^2{(m+\sqrt{m^2+k^2})^2\over m^2+k^2},\quad Q=
{m\over k}(m+\sqrt{m^2+k^2})\, ,
\labell{extrparam}
\eeq
and the interaction energy
between them, 
\beq
V_\mathrm{int}=-{m^2\over 2}{m^2+2k^2\over (m^2+k^2)^{3/2}}\,.
\eeq

On the other hand, the \textit{total} energy of the system, $E$, can be
obtained by using, \eg the formulation in \cite{HH}, with the result that
\beq
E=2m-V_\mathrm{int}\, .
\eeq 
Apparently, the sign of the interaction term is the opposite of what
might have been expected (recall that $V_\mathrm{int}<0$). This,
however, is not really the case. In isolation, the mass of each black
hole is equal to its charge. The interaction energy should be equal to
the difference between the total energy at finite separation, and the
energy when the black holes are infinitely apart, \ie the BPS mass
$2M_{BPS}=2Q$. In terms of the charge of the black holes, the parameter
$m$ is, in the approximation of
large separation, 
\beq
m=Q\left(1-{Q\over k}+O(Q/k)^2\right)\, .
\eeq
In this approximation we also have
\beq
V_\mathrm{int}=-{m^2\over k}[1+O(m/k)]\, .
\labell{vintext}\eeq
Hence we recover
\beq
E=2Q-{Q^2\over k}[1+O(Q/k)]\, ,
\eeq
\ie to this order, $E=2M_{BPS}+V_\mathrm{int}$, in which the
interaction term comes out with the expected sign.

This imposes a consistency check on the interpretation of the area as
a measure of black hole internal states. Observe that, by introducing an
interaction, the relationships $m(Q)$ and $E(Q)$ get correction terms
of order $Q/k$. Then, in principle one would expect that the entropy
$S(Q)$ should also receive corrections at that same order. However,
since $E=2M_{BPS} + V_\mathrm{int}+O(Q^3/k^2)$, the internal energy of
each black hole remains $E_\mathrm{bh}=M_{BPS}[1+O(Q^2/k^2)]$. That is,
the black holes remain in their ground state, to the order $Q/k$ we are
working. If the entropy (equal to one quarter of the horizon area) is
to be a measure of the number of internal degrees of freedom, then it
should remain the same when expressed in terms of the conserved
quantity, $Q$, that measures the number of microstates for the extremal
black hole. Hence, the corrections of order $Q/k$ to
the entropy must vanish. Indeed, from \reef{extrparam}, it follows that
this entropy is
\beq
S={A_\mathrm{bh}\over 4}=\pi Q^2[1+O(Q/k)^2]\, ,
\eeq
as required. It is therefore fully consistent to say that, despite
the interaction, and the accompanying distortion of the horizon, the
black holes are not excited above their ground state. To see the effect
of the interaction on the internal degrees of freedom of the black hole
we will have to `heat up' the system above the extremal ground state.
This we will do in the next subsection.

It should nonetheless be noted that the relation $A_\mathrm{bh}=4\pi
Q^2$ gets spoiled by corrections of order $O(Q/k)^2$. This is perhaps a
sign that the solution with the strut ceases to be sensible as the
distance between the black holes decreases\footnote{Notice, though,
that the relation $A_\mathrm{bh}=4\pi Q^2$ is neither an exact one for
an extremal black hole pierced by a cosmic string. This point deserves
further study.}. As mentioned, it becomes singular when $k=0$. In
contrast, if an external axisymmetric field is introduced so as to
exactly balance the attraction between the black holes, then the
conical singularity disappears, and the relation $A_\mathrm{bh}=4\pi
Q^2$ remains an exact one for any value of $k$ \cite{dihole}. In
addition, the solution is  non-singular (outside the horizons) even in
the limit $k=0$---the black holes reach a minimal, non-zero separation,
and do not merge. Hence, it appears that by balancing the forces
between the black holes with external fields, one obtains a system that
is physically sensible for all values of the separation.

\subsection{Microscopic entropy of the near-extremal dihole}

Just as the individual four-dimensional charged black hole solution, the
dihole can be embedded into string and M-theory in a variety of ways.
Instead of the embeddings used in \cite{fourdbh}, it is
simpler and more convenient to lift the solution to a
configuration of three M5-branes intersecting on a string, with momenta
running along the string \cite{intbrn}. If all the directions parallel to the
M5-branes, except for their common intersection, are compactified, then
one obtains the metric of a five-dimensional string, or in our case, a
pair of oppositely charged strings, with momentum running along them in
opposite directions---a `string-anti-string' pair. Let us make the
choice that $i=1,2,3$ represent the charges associated to each of the
M5-branes, and $i=4$ corresponds to the momentum along the effective
string intersection. Explicitly, the metric of the five-dimensional
string-anti-string is
\beqa
{\rm d}s^2&=&(f_1 f_2 f_3)^{1/6}\biggl[-f_4^{1/2}{\rm d}t^2+{1\over
f_4^{1/2}}\left({\rm d}y-A_{(4)}{\rm d}t\right)^2\biggr]\nonumber\\
&+&{1\over(f_1 f_2 f_3)^{1/3}}\left[
e^{(\gamma_1+\gamma_2+\gamma_3+\gamma_4)/2} ({\rm d}\rho^2+{\rm d}z^2)
+\rho^2{\rm d}\varphi^2\right]\,,
\eeqa
where $A_{(4)}$, $f_i$, $\gamma_i$ are those in \reef{tsigma} and
\reef{ai}.

It may also be useful to give the solution for the extremal limit. In
this case it is more convenient to use the coordinates and notation of
\cite{cet}, in terms of which 
\beqa
{\rm d}s^2&=&(T_1 T_2 T_3)^{1/3}\biggl[-T_4{\rm d}t^2+{1\over T_4}\bigg({\rm d}y+{2k_4
m_4\cos\theta\over \Sigma_4}\,{\rm d}t\bigg)^2\nonumber\\
&+&T_4{\Sigma_1\Sigma_2\Sigma_3\Sigma_4\over(r^2-
\kappa_+^2\cos^2\theta)^3}\bigg({{\rm d}r^2\over\Delta}+{\rm d}\theta^2\bigg)\biggr]
+{\Delta\sin^2\theta\over (T_1 T_2 T_3)^{2/3}}{\rm d}\varphi^2\,,
\eeqa
where now $\Delta=r^2-\kappa_+^2$, $\Sigma_i=(r+m_i)^2-k_i^2\cos^2\theta$,
$T_i=(\Delta +k_i^2\sin^2\theta)/\Sigma_i$. In the extremal limit, the
$f_i$ become $f_i=T_i^2$.

We will now analyze the semiclassical thermodynamics of the
four-dimensional $U(1)^4$
dihole, and then, use the effective string model to find a microscopic
explanation for the entropy. 
Again, we are interested in the situation for large distance between the
holes, \ie 
\beq
\kappa_-,\, m_i\ll\kappa_+\, .
\labell{longdist2}\eeq 
We will also take the `dilute gas'
limit for the
effective string description of the black holes near extremality, \ie 
\beq
\kappa_-,\, m_4\ll m_1,m_2,m_3\, . 
\labell{dilute}\eeq
Recall that $\kappa_-$ plays the role of a non-extremality parameter, the
extremal limit being one where $\kappa_-=0$, while the $m_i$ remain
finite.

In order to determine the thermodynamic properties of the dihole, we
compute the action for the four-dimensional solution continued to the
Euclidean section, $t\to i\tau$. The Euclidean action can be completely
reduced to a calculation of surface terms. The calculations are standard
and straightforward. The geometric parts of it yield ${\beta\over
2}\sum_i m_i-{1\over 4}A_\mathrm{tot}$, (here $A_\mathrm{tot}$ is the
sum of the areas of the horizons), plus a contribution from the strut
singularity, $-\beta V_\mathrm{int}$. The remaining contribution comes
from the gauge fields. For the dihole, the electric potentials cannot be
made to vanish simultaneously on both horizons. One can instead choose
the potential to vanish at infinity, and compute the contributions to
the action arising from the (gauge) singularities of the potential $A_0$
at the horizons. In that case, each horizon gives a contribution
$-{\beta\over 4}\sum_i Q_i\Phi_i$. Even if the charges have opposite
signs for each of the two black holes, the product $Q_i\Phi_i$ has the
same sign for both. We are interested in fixing the potentials, and not
the charges, at the horizons. For the magnetic fields, this requires the
addition of extra boundary terms to the action. Then, $\Phi_i$ is the
dual magnetic potential. With the appropriate boundary terms, the
situations with electric or magnetic charges are dual to each other, as
described in \cite{rh}. Adding up all the contributions one obtains
\beq
I=\beta\left({1\over 2}\sum_{i=1}^4 m_i-{1\over 2}\sum_{i=1}^4
Q_i\Phi_i-V_\mathrm{int}\right)
-{1\over
4}A_\mathrm{tot}\, .
\eeq

In the long distance approximation,
\beq
V_\mathrm{int}=-{1\over
4\kappa_+}\sum_{i=1}^3 Q_i^2-{m_4^2-
2\kappa_-^2\over 4\kappa_+}+O(m^3/\kappa_+^2)\, .
\labell{vintlong}\eeq
Also, since $m_i-Q_i\Phi_i=\kappa_-$, and
$A_\mathrm{tot}=4\beta\kappa_-$, we finally find
\beqa
I&=&\beta(\kappa_-- V_\mathrm{int})\nonumber\\
&=&\beta\left[\kappa_- +{1\over 4\kappa_+}\left(-2\kappa_-^2+\sum_{i=1}^4
m_i^2\right)+O(m^3/\kappa_+^2)\right]\, .
\eeqa
This action provides the grand-canonical potential
$W[T,\Phi_i]=I/\beta=E-TS_\mathrm{tot}- Q\Phi$. Here $S_\mathrm{tot}$ is
the sum of the entropy of both black holes, which we identify from their
area, and $Q\Phi=\sum_{i=1}^4{1\over 2}Q_i\Phi_i$.\footnote{Properly,
one should derive this from $W$ using standard thermodynamics. It can be
checked that, within the approximations we take, the entropy-area
relation holds. Also, one usually fixes the electric potential at
infinity, not on the horizons. It is perhaps possible to, more
rigorously, define another thermodynamic ensemble that fixes, \eg the
electric field (which is conjugate to the electric dipole moment).
Nevertheless, the present approach leads to the correct results.} The
total energy follows as
\beq
E={1\over 2}\sum_{i=1}^4 m_i+{1\over
4\kappa_+}\left(-2\kappa_-^2+\sum_{i=1}^4
m_i^2\right)+O(m^3/\kappa_+^2)\, .
\eeq

In the limit \reef{dilute}, and for large $\kappa_+$, we approximate
\beq
m_i=Q_i\left(1-{Q_i\over\kappa_+}+O(Q/\kappa_+)^2\right)\, ,\quad
i=1,\dots,3,
\labell{masscharge}\eeq
so,
\beq
E={1\over 2}\sum_{i=1}^3 Q_i-{1\over
4\kappa_+}\sum_{i=1}^3 Q_i^2+{m_4\over 2}+{m_4^2-
2\kappa_-^2\over 4\kappa_+}+O(m^3/\kappa_+^2)\, .
\labell{efirst}\eeq
Here we see that, once we have expressed the energy in terms of the
M5-brane charges $Q_{1,2,3}$, the interaction energy between the
M5-branes (the second term in the right hand side of \reef{efirst})
comes out with the expected negative sign. This is just as in the
extremal case studied above. But now we still have to account for the
pieces for the energy and interactions between the excitations along the
strings, \ie the third and fourth terms in the right hand side of
\reef{efirst}. To this effect, we have to distinguish carefully between
the energy of excitation of each string above the BPS state, call it
$\delta E$, and the interaction energy between the string-anti-string
pair, $V_\mathrm{int}$. The BPS ground state of the system consists of
the two infinitely separated M5-brane intersections, with the M5-branes
in their ground state. The energy of each intersecting brane
configuration is
$M_{BPS}={1\over 4}(Q_1+Q_2+Q_3)$. The total energy must be the sum of
three contributions,
\beq
E=2(M_{BPS}+\delta E)+V_\mathrm{int}\, .
\labell{edef}\eeq
{}From \reef{efirst}, \reef{edef}, and \reef{vintlong}, we can now 
identify
\beq
\delta E={m_4\over 4}\left(1+{m_4\over \kappa_+}\right)-{\kappa_-^2
\over 2\kappa_+}+O(m^3/\kappa_+^2)\, .
\eeq
This is the total energy carried by the lightlike left- and right-movers
along each of the two antiparallel effective strings,
\beq
\delta E=P_L+P_R\, .
\eeq 
To obtain the separate values of $P_L$ and $P_R$ we need the total
momentum along each string, $P=P_L-P_R$. This is given by the charge
associated to translations along the string direction,
\beq
P={Q_4\over
4}={\sqrt{m_4^2-\kappa_-^2}\over
4}\left(1+{m_4\over\kappa_+}+O(m^2/\kappa_+^2)\right)\, .
\eeq
At this point, it is perhaps convenient to introduce the usual boost
parameter $\delta$,
such that\footnote{Note, however, that the dihole solution with $P\neq 0$
cannot obtained by boosting the $P=0$ solution along the direction of
the string.}
\beqa
m_4&=&\kappa_-\cosh 2\delta\, ,\quad m_4+\kappa_-=2\kappa_-
\cosh^2\delta\, ,\nonumber\\
\sqrt{m_4^2-\kappa_-^2}&=&\kappa_-\sinh 2\delta\, ,\quad
m_4-\kappa_-=2\kappa_-
\sinh^2\delta\, .
\eeqa
Then, the momentum carried by left and right movers, $P_{L,R}=(\delta E\pm
P)/2$, can be expressed as
\beqa
P_L&=&{\kappa_-\over 8} e^{2\delta}\left(1+{m_4\over\kappa_+}-2{\kappa_-
\over\kappa_+}e^{-2\delta}+O(m^2/\kappa_+^2)\right)\, ,\nonumber\\
P_R&=&{\kappa_-\over 8} e^{-2\delta}\left(1+{m_4\over\kappa_+}-2{\kappa_-
\over\kappa_+}e^{2\delta}+O(m^2/\kappa_+^2)\right)\, .
\labell{lrmovers}\eeqa

With these, we can express the entropy of the dihole in terms of the
M5-brane charges and the momentum of the excitations of the string. From
\reef{composite}, we obtain, for each black
hole,
\beq
S=\pi\sqrt{m_1 m_2 m_3}\sqrt{m_4+\kappa_-}\left(1-{\kappa_-
\over\kappa_+}+{1\over 2}{\sum_{i=1}^4
m_i\over\kappa_+}+O(m^2/\kappa_+^2)\right)\, ,
\labell{entq}\eeq
the total entropy being twice this value. Putting together
\reef{masscharge}, \reef{lrmovers}, and \reef{entq},
we find that, up to the order of approximation we are considering,
\beq
S={2\pi\over\sqrt{G}}\sqrt{Q_1 Q_2 Q_3}\left(\sqrt{P_L}+\sqrt{P_R} \right)\, .
\eeq
We have reinstated here the four-dimensional Newton's constant $G$, which
until now had been set to one. The charges 
$Q_{1,2,3}$ are directly proportional to the numbers $N_{1,2,3}$
of M5-branes. The precise relationship is such that
\beq
Q_1 Q_2 Q_3={GL\over 2\pi} N_1N_2N_3\, ,
\eeq
where $L$ is the length along the effective string (the length of the
compact fifth dimension at asymptotic infinity). On the other hand,
the momentum along the string is quantized in the usual way,
\beq
P_{L,R}={2\pi\over L}N_{L,R}\, .
\eeq
Thus, the Bekenstein-Hawking entropy that we have calculated for each
black hole is
\beq
S=2\pi\sqrt{N_1N_2N_3}\left(\sqrt{N_L}+\sqrt{N_R} \right)\, .
\eeq
This is precisely the same as that obtained from the statistical mechanics of
a $1+1$ dimensional field theory with central charge $c=6N_1N_2N_3$ and
$N_{L,R}$ left- and right-moving excitations. 

Hence, the effective string model can successfully reproduce, in a
microscopic calculation, the entropy of the interacting system of a
near-extremal dihole. In string theory terms, the interaction between
the two antiparallel effective strings is due to the exchange of closed
strings between them. At large separations, this exchange is dominated
by the massless states of the closed string. These allow for a
description in terms of a classical, Coulombian and Newtonian, static
interaction potential, and the effect has been accounted for in a simple
way by a shift in the energy levels of each effective string,
\reef{edef}. At shorter distances (of order the string length) the
massive modes of the string should become relevant, and eventually a
tachyonic instability must appear. This is beyond the approximations
herewith considered. We shall simply note that, for the extremal case, a
description of the condensation of this tachyon within the supergravity
picture has been proposed in \cite{BOM}. It might be interesting to
generalize this beyond the extremal limit.

\section{Conclusions and outlook}\label{conclude}

The excitation of the black holes in the dihole above extremality has
resulted in an unexpectedly (to us) complicated solution. This is in
contrast to the individual black hole solutions, or the C-metric
solutions for pairs of black holes that accelerate apart, where the
non-extremal solutions are only slightly more involved than their
extremal counterparts. It is somewhat surprising that the extremal
solution is so much simpler, since in this case the extremal state does
not have any of the special properties, such as supersymmetry or
cancellation of forces, that are at work in other configurations.
Fortunately, near the horizons the structures simplify sufficiently so
that the physical properties of the black holes take simple, manageable
forms. 

This simplification near the horizons is also related to the reason why
the microscopic analysis in terms of an effective string does work. At
large separation, the distortion of the extremal horizons caused by the
interaction between the black holes is sufficiently small to identify
the state as a form of `near-BPS' limit. Above this state, we have
added the thermal excitations of a dilute gas of left and right movers,
and the effective string picture appears to be still reliable. We
remain in a controlled situation slightly above the supersymmetric
state, but one that goes beyond the individual near-extremal black
holes or the near-Majumdar-Papapetrou solutions of \cite{mima}.

By going beyond the extremal limit, and considering the non-extremal
diholes, we have been able to exhibit features of the microstates of the
black holes that would not have appeared otherwise. Nevertheless, in the
microscopic analysis we have needed to make a restriction to large
separation between the black holes, so as to obtain a weak strut
singularity. It is likely that a better behaved situation is obtained by
balancing the dihole (\ie cancelling the conical singularities) by
adding an external field, as was done in \cite{dihole} for the extremal
dihole. It was found there that when the dihole is exactly balanced, the
distortion of the horizons completely disappears, in the sense that they
become exact, spherically symmetric Bertotti-Robinson
throats\footnote{This is for the non-dilatonic case. In the dilatonic
cases the horizons are actually null singularities, but, when balanced,
the geometries near these singularities also become spherically
symmetric.}. Clearly, it should also be possible to similarly balance a
non-extremal dihole. Perhaps in this way one can obtain a description
which is not subject to the restriction of large separation, eqn.\
\reef{longdist}. Notice also that this possibility of balancing the
forces does not appear to be available to the equal-charge non-extremal
configurations of \cite{mima}. Unfortunately, it appears that in order
to construct the dihole solution in a background field, the magnetic
solutions need to be considered first, and the explicit form of the
potential $A_\varphi$ for the magnetic dihole is required. This we have
not been able to find for the non-extremal dihole.

Besides the interest in extending the microscopic calculations of the
entropy to more complicated systems, or in obtaining new supergravity
solutions for brane-anti-brane configurations, there is another, perhaps
more speculative motivation for studying black hole/anti-black hole
systems. In \cite{udual}, a mysterious formula for the entropy of the
generic $U(1)^4$ black hole, which includes also the neutral
Schwarzschild case, was found, which suggested an interpretation in
terms of brane-antibrane pairs. While the significance of such a formula
remains unclear, one might expect to throw some light on it by studying
black hole/anti-black hole configurations. In this respect, notice that
the coincidence limit for the non-extremal dihole is better behaved than
for the extremal one. This line remains to be pursued.

Another direction in which this work can be extended is by adding
rotation to the black holes, either parallel or antiparallel. Since the
microscopic degrees of freedom responsible for the rotation add
qualitatively new dynamics \cite{rotbh}, it would be interesting to see
how they are affected by the interaction. One can also envisage adding a
net charge to the configuration \cite{Liang}. Work on some of these
directions is currently been pursued, and will be reported elsewhere.

\section*{Noted added in proof}

Fay Dowker (informed, in turn, by David Kastor) has pointed out to us that
the extremal dihole in Einstein--Maxwell theory was previously found by
Chandrasekhar and Xanthopoulos \cite{ChXa}, albeit in a different set of
coordinates. While they correctly identified the solution as describing 
a pair of oppositely charged extremal Reissner--Nordstr\"om black holes 
joined by a string, they did not realise the equivalence of their 
solution to the Bonnor dipole. They also discussed an apparent violation 
of the extremality bound, but our analysis in Sec.~6.1 settles this
point.

It has also come to our attention that Galtsov et al.~\cite{Galtsov} also 
discovered the dilatonic generalisation of the Bonnor solution, 
independently of Davidson and Gedalin \cite{dg}. However, the dihole 
interpretation of the Bonnor dipole solution apparently remained unknown 
to them, as with other authors, until the work of \cite{dihole}.

\section*{Acknowledgments}

ET thanks Yeong Cherng Liang for useful discussions and also for his
assistance, in particular, in detecting the misprint highlighted in
Footnote 1. RE acknowledges partial support from UPV grant
063.310-EB187/98 and CICYT AEN99-0315.

\end{document}